\pdfoutput=1
\documentclass[12pt,a4paper]{cibb}

\usepackage{multirow}
\usepackage{subfigure,graphicx}
\usepackage{amsmath,latexsym,amssymb,euscript,xr}
\usepackage{booktabs}
\usepackage[nodayofweek]{datetime}
\usepackage{hyperref}
\usepackage{float}

\usepackage[table,xcdraw]{xcolor}
\usepackage{color,colortbl,tabularx}

\usepackage[english]{babel}
\usepackage[protrusion=true,expansion=true]{microtype}

\definecolor{LightBlue}{rgb}{0.88,0.9,0.9}
\definecolor{DarkBlue}{rgb}{0.74,0.76,0.76}

\definecolor{8out8}{gray}{0}
\definecolor{7out8}{gray}{0.125}
\definecolor{6out8}{gray}{0.250}
\definecolor{5out8}{gray}{0.375}
\definecolor{4out8}{gray}{0.500}
\definecolor{3out8}{gray}{0.625}
\definecolor{2out8}{gray}{0.750}
\definecolor{1out8}{gray}{0.875}

\title{\Large $\ $\\ \bf Effect of Clinical History on Predictive Model Performance for Renal Complications of Diabetes}

\author{\large Davide Dei Cas$^{1}$, Barbara Di Camillo$^{1,2}$, Gian Paolo Fadini$^{3}$, Giovanni Sparacino$^{1}$, and Enrico Longato$^{*,1}$}

\address{\footnotesize $\ $\\
$^1$ Department of Information Engineering, University of Padova, 35131 Padua, Italy \\ 
{\fontfamily{qcr}\selectfont \{deicasdavi,gianni,enrico.longato\}@dei.unipd.it, barbara.dicamillo@unipd.it} \\
$^2$ Department of Comparative Biomedicine and Food Science, University of Padova, 3513 Padua, Italy \\
$^3$ Department of Medicine, University of Padova, 35131 Padua, Italy \\
{\fontfamily{qcr}\selectfont gianpaolo.fadini@unipd.it} \\
[6pt]
$^*$corresponding author
}

\abstract{\small Diabetes, Kidney disease, Predictive modelling, Logistic regression, Clinical data. \normalsize
\\[8pt] 
{\bf Abstract.} Diabetes is a chronic disease characterised by a high risk of developing diabetic nephropathy, which, in turn, is the leading cause of end-stage chronic kidney disease. The early identification of individuals at heightened risk of such complications or their exacerbation can be of paramount importance to set a correct course of treatment.
In the present work, from the data collected in the DARWIN--Renal (DApagliflozin Real--World evIdeNce--Renal) study, a nationwide multicentre retrospective real--world study, we develop an array of logistic regression models to predict, over different prediction horizons, the crossing of clinically relevant glomerular filtration rate (eGFR) thresholds for patients with diabetes by means of variables associated with demographic, anthropometric, laboratory, pathology, and therapeutic data. In doing so, we investigate the impact of information coming from patient's past visits on the model's predictive performance, coupled with an analysis of feature importance through the Boruta algorithm.
Our models yield very good performance (AUROC as high as 0.98). We also show that the introduction of information from patient's past visits leads to improved model performance of up to 4\%. The usefulness of past information is further corroborated by a feature importance analysis.
}

\begin{document}
\thispagestyle{myheadings}
\pagestyle{myheadings}
\markright{\tt Proceedings of CIBB 2024}

\section{\bf Introduction}
\label{sec:SCIENTIFIC-BACKGROUND}

Diabetes mellitus stands as a burgeoning global health problem, ranking among the most pressing public health challenges of the 21st century. The International Diabetes Federation estimates that 537 million people were living with diabetes in 2021, with an expected increase to 783 million by the year 2045 ~\cite{IDF_atlas_2021}.
Over time, diabetes poses a significant risk for the development of various complications, both macro-- and micro--vascular. Among the latter, diabetic nephropathy emerges as the foremost cause of end--stage chronic kidney disease (CKD)~\cite{koye_diabetes_2018}.
Alarmingly, the prevalence of CKD in individuals with diabetes may be as high as double that in the general population, with diabetic nephropathy affecting approximately 40\% of patients~\cite{alicic_dkd_2017}. In light of this, identifying individuals at heightened risk of such complications or their exacerbation can be crucial to prompt comprehensive evaluation and frequent monitoring.
Consequently, predictive models capable of providing risk scores for CKD at different prediction horizons via machine learning algorithms could be valuable tools to be used as clinical decision support systems.

From a data management perspective, since diabetes is a chronic multi--factorial disease, its regular screening leads to the accumulation of a huge amount of longitudinal data, which grows rapidly over time as visits take place. 
Understanding the impact of such data on the predictive performance of models would be very important to determine which information is most relevant and which, instead, plays a secondary role.

Taking these aspects into account, in the present work, we developed machine learning models based on logistic regression to predict, over different prediction horizons, the crossing of clinically relevant glomerular filtration rate (eGFR) thresholds for patients with diabetes, whose data were collected in the DARWIN--Renal (DApagliflozin Real--World evIdeNce--Renal) study~\cite{fadini_dataset_2022}. 
In doing so, we investigated the impact of information coming from patients' past visits on the model's predictive performance, coupled with an analysis of feature importance aimed at discerning which of the variables played a primary role in the prediction. The choice to carry out these analyses by means of logistic regressions is motivated by the fact that this type of model is the most frequently used in the literature, with average AUROC values above 0.80~\cite{chen_ml_2023}.

\section{\bf Data and Outcome Definition}
\label{sec:DATA-AND-OUTCOME-DEF}

The primary source of data for the present work was the DARWIN--Renal study, a nationwide multicentre retrospective real--world study sponsored by the Italian Society of Diabetology with the support of AstraZeneca. This study was designed to examine differences in the evolution of renal function among patients with diabetes treated with dapagliflozin (a drug belonging to the sodium glucose co-transporter-2 inhibitors (SGLT2i) class) or with other non--insulin hypoglycaemic drugs.

The dataset contained records of 48,649 patients with type 2 diabetes treated at 50 diabetes specialist care clinics in Italy between 1st January 2015 and 30th September 2021 (median observation time: 2.4 years; IQR 1.1--3.8). For each subject, a number of routine follow--up visits, recorded with an irregular sampling rate, were available (median number of follow--up visits: 7.0; IQR 3.0--12.0).

\subsection{\bf \it Data Preparation}
\label{sec:DATA-PREPARATION}

Since one of the objectives of this work was to assess the impact that information from past visits might have on the predictive performance of the model, we identified a patient's baseline visit so that it had at least one visit in the past and one visit in the future relative to it. Specifically, past visits were functional in assessing the impact of past history on prediction, whereas future visits allowed to define the occurrence or non-occurrence of the renal outcomes of interest. Given the irregular sampling rate between visits, and especially between the first two, we performed this subdivision, for each subject, according to the following pipeline:
\begin{enumerate}
    \itemsep0pt
    \item Considering the 9 months since the date of the first follow-up visit, the most recent visit, i.e., the last one within these 9 months, was designated as the baseline.
    \item From the date of the baseline, all visits that fell within the preceding 9 months were regarded as past visits, whereas other visits prior to 9 months were excluded because too remote relative to the baseline.
    \item All visits after the baseline, with the exception of those with missing eGFR, were considered future visits and only used to check for the occurrence of the renal outcomes of interest.
\end{enumerate}
In practice, this procedure allowed to homogenise the longitudinal information on each subject and reduce any possible bias due to the trial-like data collection procedure of the original study.

After excluding subjects with insufficient future information to define the outcomes, the total sample size was 32,379.
Each subject was characterised by 28 variables associated with demographic, anthropometric, laboratory, pathology, and therapeutic data.
To provide the model with information from a patient's past visits, we chained to the 28 variables of the baseline, 6 new variables representing the average on past visits of weight, diastolic blood pressure (DBP), systolic blood pressure (SBP), glycated haemoglobin (HbA1c), eGFR, and albumin excretion rate (AER).
The complete list of 34 variables used in the analysis, with their description and characterisation, is shown in~\autoref{tab:VARIABLES}.

\begin{table}[H]
\centering
\caption{\textbf{Input variables}. Variables used in the analysis with extended description, availability percentage and population description at the baseline visit. Continuous variables are described by the mean $\pm$ standard deviation, dichotomous variables as the number of 1s (percentage relative to the number of subjects).
\textit{Note}: calculations were made ignoring subjects with missing values for the specific variable.}
\label{tab:VARIABLES}
\resizebox{0.60\textwidth}{!}{%
\begin{tabular}{@{}llcc@{}}
\toprule
\textbf{Variable name} & \textbf{Description} & \textbf{Available \%} & \textbf{\begin{tabular}[c]{@{}c@{}}Population characteristics\\ (N = 32,379)\end{tabular}} \\ \midrule
\rowcolor{LightBlue} Sex & Male sex & 100 & 19574 (60.5\%) \\
Age & Age {[}years{]} & 100 & 63.6 $\pm$ 8.7 \\
\rowcolor{LightBlue} Duration & Diabetes duration {[}years{]} & 98.7 & 12.1 $\pm$ 8.1 \\
Weight & Weight {[}Kg{]} & 81.6 & 84.2 $\pm$ 17.7 \\
\rowcolor{LightBlue} BMI & Body mass index {[}Kg/m\textsuperscript{2}{]} & 77.6 & 30.3 $\pm$ 5.7 \\
SBP & Systolic blood pressure {[}mmHg{]} & 65.4 & 134.9 $\pm$ 18.1 \\
\rowcolor{LightBlue} DBP & Diastolic blood pressure {[}mmHg{]} & 65.3 & 77.4 $\pm$ 9.6 \\
HbA1c & Glycated haemoglobin {[}\%{]} & 94.3 & 7.4 $\pm$ 1.1 \\
\rowcolor{LightBlue} eGFR & Estimated glomerular filtration rate  {[}mL/min/1.73 m\textsuperscript{2}{]} & 72.1 & 79.4 $\pm$ 21.4 \\
AER & Albumin excretion rate {[}mg/g{]} & 45.2 & 96.9 $\pm$ 464.5 \\
\rowcolor{LightBlue} CKD & Chronic kidney disease & 46.3 & 8065 (53.8\%) \\
ACR\textgreater30 & Pathologic albuminuria & 45.2 & 4478 (30.6\%) \\
\rowcolor{LightBlue} METF & Metformin & 100 & 26328 (81.3\%) \\
SU/Rep & Sulphonylurea / Repaglinide & 100 & 8647 (26.7\%) \\
\rowcolor{LightBlue} DPP4i & DPP--4 inhibitors & 100 & 11089 (34.2\%) \\
GLP1RA & GLP--1 receptor agonists & 100 & 5343 (16.5\%) \\
\rowcolor{LightBlue} SGLT2i & SGLT--2 inhibitors & 100 & 9983 (30.8\%) \\
PIOGLIT & Pioglitazone & 100 & 2827 (8.7\%) \\
\rowcolor{LightBlue} ACARB & Acarbose & 100 & 1152 (3.6\%) \\
BOLUS & Bolus insulin & 100 & 4846 (15.0\%) \\
\rowcolor{LightBlue} BASAL & Basal insulin & 100 & 10570 (32.6\%) \\
STATIN & Statins & 100 & 18223 (56.3\%) \\
\rowcolor{LightBlue} APA & Anti--platelet agents & 100 & 12973 (40.1\%) \\
ACEi/ARB & ACE inhibitors / Angiotensin receptor blockers & 100 & 19031 (58.8\%) \\
\rowcolor{LightBlue} BBLock & $\beta$-blockers & 100 & 9197 (28.4\%) \\
CCB & Calcium channel inhibitors & 100 & 6936 (21.4\%) \\
\rowcolor{LightBlue} Diuretics & Diuretics & 100 & 10000 (30.9\%) \\
Acoag & Anticoagulants & 100 & 956 (3.0\%) \\
\rowcolor{LightBlue} Weight\_past & Average weight on past visits (9 months) {[}Kg{]} & 89.0 & 84.9 $\pm$ 17.9 \\
SBP\_past & Average SBP on past visits (9 months) {[}mmHg{]} & 73.1 & 135.5 $\pm$ 17.5 \\
\rowcolor{LightBlue} DBP\_past & Average DBP on past visits (9 months) {[}mmHg{]} & 73.0 & 77.9 $\pm$ 9.3 \\
HbA1c\_past & Average HbA1c on past visits (9 months) {[}\%{]} & 98.3 & 7.6 $\pm$ 1.2 \\
\rowcolor{LightBlue} eGFR\_past & Average eGFR on past visits (9 months) {[}mL/min/1.73 m\textsuperscript{2}{]} & 83.6 & 80.1 $\pm$ 21.1 \\
AER\_past & Average AER on past visits (9 months) {[}mg/g{]} & 60.0 & 106.7 $\pm$ 581.6 \\ \bottomrule
\end{tabular}%
}
\end{table}

\subsection{\bf \it Outcome Definition}
\label{sec:OUTCOME-DEFINITION}

To describe renal function deterioration, we considered 5 soft outcomes defined as the crossing of the 6 renal disease severity thresholds established by the KDOQI (Kidney Disease Outcomes Quality Initiative) scale~\cite{nkf_kdoqi_2002} based on the eGFR value. Specifically, the thresholds were 90, 60, 45, 30, and 15 mL/min/1.73 m\textsuperscript{2}, i.e., those between stages I, II, IIIa, IIIb, IV, and V, respectively.
We predicted these threshold crossings at eight prediction horizons, namely 6, 12, 18, 24, 30, 36, 42, and 48 months.

\section{\bf Model Development}
\label{sec:MODEL-DEVELOP}

For each combination of outcome and prediction horizon, we trained a logistic regression model. For this purpose, we split the data into training and test sets following a 80\%--20\% split. We performed hyperparameters tuning by means of a 5--fold cross--validation using the same 5 folds from the training set for each model. The tested hyperparameters were the type of regularisation (L1 or L2) and the regularisation coefficient C (grid search in the set \{0.001, 0.01, 0.1, 1, 5, 10, 50, 100, 500\}). The input of each model was the baseline visit of each patient described by the variables in~\autoref{tab:VARIABLES}, initially excluding the last 6 dynamic variables in order to evaluate the models without the introduction of information associated with past visits. Missing values (NA) were imputed using the MICE (Multiple Imputation by Chained Equations) method~\cite{azur_mice_2011}. The output of each model was instead represented by the estimate of the probability that a patient had of developing the specific kidney disease outcome at the considered prediction horizon. Subjects with unknown outcomes at the specific time horizon were excluded from that part of the analysis. We evaluated the predictive performance of each model on the same test set, using the area under the receiver-operating characteristic curve (AUROC).

Next, for each outcome and prediction horizon, we evaluated the impact of introducing information from a patient's clinical history by considering also the 6 variables representing the average on past visits of weight, DBP, SBP, HbA1c, eGFR, and AER. This evaluation reflected the first aim of this work, namely to test whether the inclusion of information from past visits, here encoded by the mean, together with data collected at the baseline visit, could increase the predictive power of the models. Intuitively, this could happen because the model would then have at its disposal not only a single point value, but also a description of the first order with which to compare it (e.g., if the weight at the baseline visit was lower than the past average, this means that the patient was losing weight).

We also investigated which variables played a primary role in prediction and which, instead, had a lower impact. To perform this feature importance analysis, we applied the Boruta algorithm~\cite{kursa_boruta_2010}. By repeating the training of all logistic regression models, including the 6 variables representative of the past, we performed the selection of the most important variables. In contrast to previous models, in this case, the only evaluated hyperparameter was the regularisation coefficient C, while the regularisation type was fixed at L2 (as L1 would uncontrollably remove additional features).

\section{\bf Results}
\label{sec:RESULTS}

~\autoref{tab:PERFORMANCE} presents the AUROC values, for each combination of renal function outcome and prediction horizon, obtained using: A) basic models without the introduction of past history and without feature selection, B) models with the introduction of the 6 variables related to past visits, and C) models with the additional use of feature selection. 

\begin{table}[H]
\centering
\caption{\textbf{Model performance}. Comparison of the performance, in terms of AUROC, of models without the introduction of past history and without feature selection (H: No - FS: No) with models with the introduction of past history but without feature selection (H: Yes - FS: No) and with models with the introduction of past history and feature selection (H: Yes - FS: Yes) for all 5 eGFR threshold crossings.
\textit{Abbreviations}: H, history; FS, feature selection.
\textit{Note}: statistical significance vs. (H: No - FS: No) model marked with *. Statistical significance vs. (H: Yes - FS: No) model marked with \dag.}
\label{tab:PERFORMANCE}
\resizebox{0.55\textwidth}{!}{%
\begin{tabular}{@{}clllllllll@{}}
\toprule
\multirow{2}{*}{\textbf{\begin{tabular}[c]{@{}c@{}}Outcome\\ $[$mL/min/1.73 m\textsuperscript{2}$]$\end{tabular}}} & \multicolumn{1}{c}{\multirow{2}{*}{\textbf{Model}}} & \multicolumn{8}{c}{\textbf{Prediction horizon $[$months$]$}} \\
 & \multicolumn{1}{c}{} & \multicolumn{1}{c}{\textbf{6}} & \multicolumn{1}{c}{\textbf{12}} & \multicolumn{1}{c}{\textbf{18}} & \multicolumn{1}{c}{\textbf{24}} & \multicolumn{1}{c}{\textbf{30}} & \multicolumn{1}{c}{\textbf{36}} & \multicolumn{1}{c}{\textbf{42}} & \multicolumn{1}{c}{\textbf{48}} \\ \midrule
\multirow{3}{*}{\textbf{90}} & \textbf{H: No - FS: No} & 0.747 & 0.776 & 0.772 & 0.770 & 0.769 & 0.786 & 0.784 & 0.805 \\
 & \textbf{H: Yes - FS: No} & 0.780* & 0.791* & 0.788* & 0.787* & 0.783* & 0.798 & 0.797 & 0.827 \\
 &  \textbf{H: Yes - FS: Yes} & 0.801*\dag & 0.797* & 0.793* & 0.789* & 0.788* & 0.797 & 0.793 & 0.830 \\ \midrule
\multirow{3}{*}{\textbf{60}} & \textbf{H: No - FS: No} & 0.858 & 0.870 & 0.883 & 0.882 & 0.869 & 0.872 & 0.865 & 0.875 \\
 &  \textbf{H: Yes - FS: No} & 0.883* & 0.885* & 0.895* & 0.891* & 0.879* & 0.881* & 0.875* & 0.888* \\
 & \textbf{H: Yes - FS: Yes} & 0.884* & 0.885* & 0.895* & 0.890* & 0.879* & 0.880* & 0.872 & 0.886 \\ \midrule
\multirow{3}{*}{\textbf{45}} &  \textbf{H: No - FS: No} & 0.919 & 0.908 & 0.915 & 0.911 & 0.906 & 0.900 & 0.895 & 0.881 \\
 & \textbf{H: Yes - FS: No} & 0.919 & 0.917* & 0.922* & 0.915 & 0.910 & 0.906* & 0.901* & 0.886 \\
 &  \textbf{H: Yes - FS: Yes} & 0.922 & 0.918* & 0.922* & 0.914 & 0.908 & 0.903 & 0.899 & 0.882 \\ \midrule
\multirow{3}{*}{\textbf{30}} & \textbf{H: No - FS: No} & 0.947 & 0.885 & 0.913 & 0.921 & 0.909 & 0.899 & 0.900 & 0.893 \\
 & \textbf{H: Yes - FS: No} & 0.957* & 0.892 & 0.922* & 0.926* & 0.911 & 0.906* & 0.905 & 0.902* \\
 & \textbf{H: Yes - FS: Yes} & 0.957* & 0.902*\dag & 0.928* & 0.927 & 0.916 & 0.903 & 0.902 & 0.901 \\ \midrule
\multirow{3}{*}{\textbf{15}} &  \textbf{H: No - FS: No} & 0.984 & 0.881 & 0.902 & 0.903 & 0.912 & 0.903 & 0.903 & 0.884 \\
 & \textbf{H: Yes - FS: No} & 0.983 & 0.876 & 0.893 & 0.895 & 0.905 & 0.900 & 0.901 & 0.883 \\
 & \textbf{H: Yes - FS: Yes} & 0.972 & 0.888 & 0.898 & 0.904 & 0.909 & 0.901 & 0.903 & 0.890 \\ \bottomrule
\end{tabular}%
}
\end{table}

The predictive performance of all models was satisfactory for all outcomes and for all prediction horizons, with values never below 0.74.
The outcome that appeared most complex to predict was the 90 mL/min/1.73 m\textsuperscript{2} threshold crossing, where AUROC values were often below 0.80. For all other outcomes, performances were always above 0.85, even reaching values above 0.90, especially for predictions made within 24 months, thus improving the performance of most models found in the literature.

The results showed a clear improvement in performance (from about 1 to 4\%) with the introduction of past visit information into the model. Except for the 15 mL/min/1.73 m\textsuperscript{2} threshold crossing, where there appeared to be no statistically significant differences, for all other outcomes, statistical significance of the improvement was detectable at almost all prediction horizons. The absence of statistical significance in the case of 15 mL/min/1.73 m\textsuperscript{2} threshold crossing could be attributed to the fact that the number of subjects showing this outcome within the dataset was small (even after 48 months, the number of subjects for whom this outcome was observed was $<$60).

The introduction of feature selection resulted in models with performance at least as good as those without it, with the exception of two cases (90 mL/min/1.73 m\textsuperscript{2} crossing at 6 months and 30 mL/min/1.73 m\textsuperscript{2} crossing at 12 months), where a statistically significant improvement was observed due to the introduction of feature selection.
The variables that were identified by the Boruta algorithm as most significant are shown, for each clinical outcome, in~\autoref{tab:SELECTED-VARIAB}. 
From this feature importance analysis, it emerged that, of the 6 dynamic variables representative of past visits, eGFR and AER values prior to the baseline played an important role in prediction since they were chosen, respectively, in 100\% and 92.5\% of the cases, on average. In contrast, past DBP and past HbA1c values seemed to have a minor impact, being selected less frequently. Finally, past values of SBP and weight never appeared to play particularly important role.
As for the other variables, those that appeared most useful in predicting all the clinical outcomes assessed were age, eGFR, and AER at the baseline. For the outcomes related to the 90, 60 and 45 mL/min/1.73 m\textsuperscript{2} threshold crossings, the duration of diabetes also seemed to be important, while for the 45, 30 and 15 mL/min/1.73 m\textsuperscript{2} thresholds crossing, the variables CKD and ACR\textgreater30 were important. The glucose lowering drugs that were most frequently selected were metformin and SGLT2i, two drugs for which negative and positive effects on renal disease, respectively, have been highlighted~\cite{hsu_metf_2018, fadini_sglt2i_2024}, while, with regard to the other classes of drugs, diuretics, $\beta$--blockers, and ACEi/ARB appeared important.

\begin{table}[H]
\centering
\caption{\textbf{Feature importance}. Fraction and percentage of prediction horizons at which each variable is selected by the Boruta algorithm for the considered clinical outcomes. 
\textit{Note}: blank spaces indicates that the variable was never selected for the outcome, darker shade of grey indicates a higher selection percentage.}
\label{tab:SELECTED-VARIAB}
\resizebox{0.51\textwidth}{!}{%
\begin{tabular}{@{}lccccc@{}}
\toprule
 & \multicolumn{5}{c}{\textbf{Outcome $[$mL/min/1.73 m\textsuperscript{2}$]$}} \\
\multirow{-2}{*}{\textbf{Variables}} & \textbf{90} & \textbf{60} & \textbf{45} & \textbf{30} & \textbf{15} \\ \midrule
\textbf{Sex} &  &  &  &  &  \\
\textbf{Age} & \cellcolor{8out8}\textcolor{white}{8/8 (100\%)} & \cellcolor{8out8}\textcolor{white}{8/8 (100\%)} & \cellcolor{8out8}\textcolor{white}{8/8 (100\%)} & \cellcolor{8out8}\textcolor{white}{8/8 (100\%)} & \cellcolor{3out8} 3/8 (37.5\%) \\
\textbf{Duration} & \cellcolor{4out8}4/8 (50\%) & \cellcolor{8out8}\textcolor{white}{8/8 (100\%)} & \cellcolor{8out8}\textcolor{white}{8/8 (100\%)} &  &  \\
\textbf{Weight} &  &  &  &  &  \\
\textbf{BMI} &  & \cellcolor{3out8} 3/8 (37.5\%) &  &  &  \\
\textbf{SBP} &  & \cellcolor{7out8}\textcolor{white}{7/8 (87.5\%)} &  &  &  \\
\textbf{DBP} &  &  &  &  &  \\
\textbf{HbA1c} &  &  & \cellcolor{1out8} 1/8 (12.5\%) &  &  \\
\textbf{eGFR} & \cellcolor{8out8}\textcolor{white}{8/8 (100\%)} & \cellcolor{8out8}\textcolor{white}{8/8 (100\%)} & \cellcolor{8out8}\textcolor{white}{8/8 (100\%)} & \cellcolor{8out8}\textcolor{white}{8/8 (100\%)} & \cellcolor{8out8}\textcolor{white}{8/8 (100\%)} \\
\textbf{AER} & \cellcolor{8out8}\textcolor{white}{8/8 (100\%)} & \cellcolor{8out8}\textcolor{white}{8/8 (100\%)} & \cellcolor{8out8}\textcolor{white}{8/8 (100\%)} & \cellcolor{8out8}\textcolor{white}{8/8 (100\%)} & \cellcolor{8out8}\textcolor{white}{8/8 (100\%)} \\
\textbf{CKD} &  & \cellcolor{1out8} 1/8 (12.5\%) & \cellcolor{8out8}\textcolor{white}{8/8 (100\%)} & \cellcolor{8out8}\textcolor{white}{8/8 (100\%)} & \cellcolor{8out8}\textcolor{white}{8/8 (100\%)} \\
\textbf{ACR$>$30} &  &  & \cellcolor{8out8}\textcolor{white}{8/8 (100\%)} & \cellcolor{8out8}\textcolor{white}{8/8 (100\%)} & \cellcolor{7out8}\textcolor{white}{7/8 (87.5\%)} \\
\textbf{METF} &  &  & \cellcolor{7out8}\textcolor{white}{7/8 (87.5\%)} & \cellcolor{8out8}\textcolor{white}{8/8 (100\%)} & \cellcolor{8out8}\textcolor{white}{8/8 (100\%)} \\
\textbf{SU/Rep} &  &  &  &  &  \\
\textbf{DPP4i} &  &  &  &  &  \\
\textbf{GLP1RA} &  &  &  &  &  \\
\textbf{SGLT2i} & \cellcolor{1out8} 1/8 (12.5\%) &  & \cellcolor{2out8} 2/8 (25\%) & \cellcolor{3out8} 3/8 (37.5\%) &  \\
\textbf{PIOGLIT} &  &  &  &  &  \\
\textbf{ACARB} &  &  &  &  &  \\
\textbf{BOLUS} &  &  &  &  &  \\
\textbf{BASAL} &  &  &  &  &  \\
\textbf{STATIN} &  &  &  &  &  \\
\textbf{APA} & \cellcolor{1out8} 1/8 (12.5\%) & \cellcolor{3out8} 3/8 (37.5\%) &  &  &  \\
\textbf{ACEi/ARB} &  & \cellcolor{5out8}\textcolor{white}{5/8 (62.5\%)} & \cellcolor{4out8}4/8 (50\%) &  &  \\
\textbf{BBLock} &  & \cellcolor{2out8} 2/8 (25\%) & \cellcolor{4out8}4/8 (50\%) & \cellcolor{3out8} 3/8 (37.5\%) &  \\
\textbf{CCB} &  &  &  &  &  \\
\textbf{Diuretics} &  & \cellcolor{7out8}\textcolor{white}{7/8 (87.5\%)} & \cellcolor{8out8}\textcolor{white}{8/8 (100\%)} & \cellcolor{7out8}\textcolor{white}{7/8 (87.5\%)} &  \\
\textbf{Acoag} &  &  &  &  &  \\
\textbf{Weight\_past} &  &  &  &  &  \\
\textbf{DBP\_past} &  & \cellcolor{5out8}\textcolor{white}{5/8 (62.5\%)} & \cellcolor{5out8}\textcolor{white}{5/8 (62.5\%)} & \cellcolor{3out8} 3/8 (37.5\%) &  \\
\textbf{SBP\_past} &  &  &  &  & \cellcolor{1out8} 1/8 (12.5\%) \\
\textbf{HbA1c\_past} & \cellcolor{1out8} 1/8 (12.5\%) & \cellcolor{3out8} 3/8 (37.5\%) & \cellcolor{2out8} 2/8 (25\%) &  & \cellcolor{4out8}4/8 (50\%) \\
\textbf{eGFR\_past} & \cellcolor{8out8}\textcolor{white}{8/8 (100\%)} & \cellcolor{8out8}\textcolor{white}{8/8 (100\%)} & \cellcolor{8out8}\textcolor{white}{8/8 (100\%)} & \cellcolor{8out8}\textcolor{white}{8/8 (100\%)} & \cellcolor{8out8}\textcolor{white}{8/8 (100\%)} \\
\textbf{AER\_past} & \cellcolor{5out8}\textcolor{white}{5/8 (62.5\%)} & \cellcolor{8out8}\textcolor{white}{8/8 (100\%)} & \cellcolor{8out8}\textcolor{white}{8/8 (100\%)} & \cellcolor{8out8}\textcolor{white}{8/8 (100\%)} & \cellcolor{8out8}\textcolor{white}{8/8 (100\%)} \\ \bottomrule
\end{tabular}%
}
\end{table}

\section{\bf Discussion and Conclusions}
\label{sec:CONCLUSIONS}

This work investigated the influence of a patient's clinical history on the performance of predictive models of diabetes' renal complications by introducing 6 variables representative of values collected during past visits.
The obtained performance was satisfactory, with models showing AUROC values, on average, well above 0.8 and a marked improvement in performance associated with the introduction of information from the past. In addition, feature importance analysis further highlighted the usefulness of introducing these variables (especially past eGFR and past AER). 

Future developments may include: testing a wider range of modelling approaches (e.g., deep learning models), and a different management of past visits, in order to provide the model not only with an average value representing the past but with the entire sequence of values.

\section*{\bf Conflict of interests}
\label{sec:CONFLICT-OF-INTERESTS}
GPF received grant support, honoraria or lecture fees from Abbott, AstraZeneca, Boehringer, Lilly, MSD, Novartis, Novo Nordisk, Sanofi, Takeda, Servier. DDC, BDC, GS, and EL have nothing to disclose.

\section*{\bf Acknowledgments}
\label{sec:ACKNOWLEDGMENTS}
We wish to thank the Italian Diabetes Society for providing access to the DARWIN-Renal database.

\section*{\bf Funding}
\label{sec:FUNDING}
This work was supported by the Complementary National Plan PNC--I.1 “Research initiatives for innovative technologies and pathways in the health and welfare sector” D.D. 931 of 06/06/2022, DARE -- DigitAl lifelong pRevEntion initiative, code PNC0000002, CUP: B53C2200644000.

\footnotesize
\bibliographystyle{unsrt}
\bibliography{main.bib} 
\normalsize

\end{document}